\documentclass[twocolumn,aps,10pt,longbibliography,showkeys]{revtex4-2}
\usepackage{amsmath}
\usepackage{bm}
\usepackage{amssymb,latexsym,mathrsfs}

\usepackage{url}
\usepackage{color}

\usepackage{tikz}
\usepackage{xcolor}
\usepackage[bookmarks=false,breaklinks]{hyperref}

\definecolor{royalbluecs}{rgb}{0.2, 0.2, 1}

\hypersetup{linktoc=all, 
    colorlinks=true, linkcolor={royalbluecs},
    citecolor={royalbluecs}, urlcolor={royalbluecs}}

\definecolor{lime}{HTML}{A6CE39}
\DeclareRobustCommand{\orcidicon}{%
	\begin{tikzpicture}
	\draw[lime, fill=lime] (0,0) 
	circle [radius=0.16] 
	node[white] {{\fontfamily{qag}\selectfont \tiny ID}};
	\draw[white, fill=white] (-0.0625,0.095) 
	circle [radius=0.007];
	\end{tikzpicture}
	\hspace{-2mm}
}

\foreach \x in {A, ..., Z}{%
	\expandafter\xdef\csname orcid\x\endcsname{\noexpand\href{https://orcid.org/\csname orcidauthor\x\endcsname}{\noexpand\orcidicon}}
}

\long\def\comment#1{} 

\normalsize
 
\begin{document}

\title{From maximum force to physics in 9 lines and towards relativistic quantum gravity}

\author{Christoph Schiller \orcidA{}}
\email{cs@motionmountain.net}
\affiliation{Motion Mountain Research, 81827 Munich, Germany}

\date{7 November 2022} %

\begin{abstract} 
\noindent A compact summary of present fundamental physics is given and evaluated.  Its 9 lines describe all observations exactly and contain both general relativity and the standard model of particle physics.  Their precise agreement with experiments, in combination with their extreme simplicity and their internal consistency, suggest that there are no experimental effects beyond the two theories.  The 9 lines imply a smallest length in nature and make concrete suggestions for the microscopic constituents in a complete theory of relativistic quantum gravity.  It is shown that the microscopic constituents cannot be described by a Lagrangian or by an equation of motion. Finally, the 9 lines specify the only decisive tests that allow checking any specific proposal for such a theory.
\end{abstract}



\maketitle



\noindent 
In fundamental physics, a world-wide search for the theory of relativistic quantum gravity is under way.
Despite intense attempts in experiment and theory, the search is still ongoing.
So far, all experiments ever performed and all observations ever made can be described with general relativity and with the standard model of particle physics (with massive Dirac neutrinos, as always implied in this article \cite{wells2020once}). %
The world-wide search for observations beyond general relativity was unsuccessful \cite{will,gtc3,doublepulsar}, and so was the world-wide search for observations beyond the standard model \cite{pdgnew}.

{\small
\begin{table}[b!]  
\def\boldsymbol#1{#1}\relax
\let\pmb\relax
\let\textbf\relax
\setlength{\tabcolsep}{0.5mm}%
\caption{Nine lines describe all observations about nature.\hss}
\vspace*{1mm}
\begin{tabular*}{0.48\textwidth}{%
@{\hspace{0em}}p{6mm}@{\hspace{0em}}%
@{\extracolsep{\fill}} p{20mm}@{\hspace{0em}}%
@{\extracolsep{\fill}} p{58mm}@{\hspace{0em}}}
\hline
{Nr.} & {Line} & {Details}\\
\hline
%
%
(1) & 	$\boldsymbol{{\rm d}} \boldsymbol{W} \boldsymbol{=} \boldsymbol{0}$ 	        &   Action $W=\int L \,{\rm d}t$ is minimized in local motion. The lines below fix the two fundamental Lagrangians $L$.\\
(2) &   $\boldsymbol{v} \boldsymbol{\leq} \boldsymbol{c} $ 	        &   Energy speed $v$ is limited by the speed of light $c$. This invariant implies special relativity and restricts the possible Lagrangians.\\
(3) & 	$\boldsymbol{F} \boldsymbol{\leq}  \boldsymbol{c^4/4G}$    &	Force $F$ is limited by $c$ and by the gravitational constant $G$. This invariant implies general relativity and, together with lines 1 and 2, \textit{fixes} its Hilbert Lagrangian.\\
(4) &   $\boldsymbol{W} \boldsymbol{\geq}  {\pmb{\hbar}} $ 	&   Action $W$ is never smaller than the quantum of action  $ \hbar $. This invariant implies quantum theory and restricts the possible Lagrangians.\\
(5) & 	$\boldsymbol{S} \boldsymbol{\geq} \boldsymbol{k} \, {\boldsymbol{\ln}} \boldsymbol{2}$ &   Entropy $S$ is never smaller than 
$\ln 2$
times the Boltzmann constant $k$. This invariant implies thermodynamics.\\
(6) &   \textbf{U(1)} 	            &   is the gauge group of the
                                      electromagnetic interaction. It yields
                                      its Lagrangian when combined with lines 1, 2 and 4.\\
(7) &   \hbox{\textbf{SU(3)~and}} \hbox{\textbf{broken SU(2)\hfil}} &   
                                        are the gauge groups of the two
                                        nuclear interactions, yielding their
                                        Lagrangians when combined with lines 1, 2 and 4.\\
(8) &   \textbf{18~particles}\hfil  &   – gauge bosons, the Higgs boson, quarks, leptons, and the undetected graviton – with all their quantum numbers \cite{pdgnew}, make up everything and, with the interactions, \textit{fix} the standard model Lagrangian.\\
(9) &   \hbox{\textbf{Finally,~27}} \hbox{\textbf{numbers}}\hfil   &   – dimensions, cosmological constant, coupling constants, particle mass ratios, mixing matrices \cite{pdgnew} – \textit{complete} the two fundamental Lagrangians. They determine all observations, including all colours.\\
\hline
\end{tabular*}
\end{table}}

The present article first summarizes general relativity and the standard model in a way that is as simple and as compact as possible, while keeping the precision that the two theories provide. %
The summary consists of 9 short lines, each of them decades old: 
five general principles and four lines of specific choices.
Evaluating the 9-line summary shows that it contains all equations of fundamental physics. 
The 9~lines also highlight the open issues in the foundations of physics.
The simplicity of the summary yields explicit experimental predictions.
The 9~lines also define the requirements that any theory of relativistic quantum gravity must fulfil, in particular about the type and behaviour of the microscopic constituents of nature.
In particular, the 9~lines lead to a limited number of decisive tests. 
The deduced requirements and tests explain why relativistic quantum gravity has not yet been achieved, and provide guidance for future searches.

\section{Least action}


\noindent 
In nature, all motion can be described by the principle of least action: \emph{motion minimizes action.}
More precisely, this applies to microscopic motion; 
on large scales, action can also be stationary.

In everyday life, action is the time integral of the Lagrangian, i.e., of the difference between kinetic and potential energy  \cite{landaul9l}. %
In the general case, the action is defined as the integral of a Lagrangian density based on and built with observable fields.
The equations of motion follow from the requirement that action is minimized -- or stationary. 

The history of the principle of least action is complicated and long. 
After the first precise description of motion by Galileo, researchers took about 150 years to 
complete the definition of `action'.  %
In physics, action $W$ is a scalar quantity measuring the change occurring in a system.
Measurement of action is based on the ability to measure length and time intervals. 

Experimental validation of the principle of least (or stationary) action 
occurs every day -- in classical physics, in quantum theory and in general relativity. 
Action minimization describes every type of motion. 
Action minimization is valid for the motion of machines, molecules, animals, electricity and light, 
for the motion of planets and stars, for the motion of particles and fields, and for the change of curvature of empty space. %
(\textcolor{black}{Also the path integral formulation of quantum field theory can be taken as following from a Lagrangian.})
Falsification of least (or stationary) action requires finding an exception in an observation. 
In principle, this is possible, but the probability is low.
In fact, no non-equivalent alternative to the principle of least (or 
stationary) action appears to have ever been proposed.

\textcolor{black}{It is important to note that the principle of least (or stationary) action also implies and contains the principle of observer invariance, because the action for an evolving system is defined to have the same value for all observers. 
Depending on the set of observers being studied, the invariance 
and the underlying symmetry differs, as shown in the remaining lines.}

In short, on a small scale, all motion follows the \emph{principle of 
least (or stationary) action} 
\begin{equation}
	{\rm d}W=0 \;\;.
\end{equation}
Microscopic motion minimizes action $W=\int L \,{\rm d}t$, i.e., minimizes the integral of the Lagrangian $L$.
The two fundamental Lagrangians of nature, the Hilbert Lagrangian of general relativity and 
the Lagrangian of the standard model of particle physics, are defined in the following. %








\section{Maximum speed}


\noindent 
Special relativity is based on the principle of an invariant maximum speed with a 
value $c \approx 3.0 \cdot 10^{8} \rm \,m/s$. 
In nature, energy cannot move faster than $c$.
The maximum speed itself is only achieved by massless radiation, such as electromagnetic 
or gravitational waves. 
Maximum speed is the origin of the Lorentz transformations, the mixing of space and time, 
the equivalence of energy and mass, the relativity of time, the relativity of length, and the speed addition formula. %
The invariant limit property of $c$ thus goes beyond a conversion factor between length and time.

Maximum speed was discovered in the years from 1860 to 1890.
In 1905, Einstein deduced the Lorentz transformations from maximum speed \cite{einsteinc}.
In particular, maximum speed $c$ determines the form of any Lagrangian that complies with special relativity.
In particular, because action is observer-invariant, it must be a Lorentz scalar.

In a vacuum, light from a moving lamp has the same speed as light from a lamp at rest.
Experimentally, this holds in all directions \cite{Antonini:2005yb}.
Furthermore, even the comparatively light electrons cannot be accelerated faster than light, even using the largest amounts of energy.  %
This speed limit is found to apply also to protons, neutrinos, rockets, radio waves, X-rays and gravitational waves.
The speed limit is so fundamental that it is used to define the meter as the path of light during a given interval of time. %
No type of matter and no type of radiation moves faster than $c$.
The speed limit is a \emph{local} limit: it is valid for energy speeds at a single point.
Sums of speeds at different locations can exceed the limit. 
This aspect is of importance in the next section.

Experimental validation of maximum speed is frequent. 
Every electric motor confirms the existence of a maximum speed.
No known example of motion of energy contradicts maximum speed. 
Maximum speed is valid in classical physics, in quantum theory and in general relativity. 
Falsification means finding a system in which energy moves faster than $c$. 
Such an observation is possible in principle, but the probability is low.
Despite high potential rewards, nobody has found a way to move energy faster than light in vacuum.
Likewise, attempts to find a description of nature without maximum energy speed have not been successful.

\textcolor{black}{Again, it is important to note that the principle of maximum speed, together with the principle of least (or stationary) action, implies and contains Lorentz invariance.}

In short, special relativity can be deduced 
\textcolor{black}{from the 
principle of least (or stationary action) together}
with the \emph{principle of maximum speed}
\begin{equation}
	v \leq c \;\;.
\end{equation}
There is an energy speed limit in nature.
Among others, the principle requires that Lagrangians must be Lorentz-scalars.

\section{Maximum force}


\noindent 
In 1973, Elizabeth Rauscher discovered that general relativity \emph{implies} a limit to force: she assumed that is was given by the 
force $F=c^4/G$ \cite{rauscher}. %
She was followed by many other researchers \cite{treder,10.2307/24530850,sab,massa,Kostro:1999ue,gibbons,s1,csmax,csmax2,barrow1,abo,ong2,Bolotin:2016roy,maxlum,Ong:2018xna,barrow2,barrow3,b1,atazadeh,jowsey,faraoni,newcs,Faraoni:2021sep,MYNEWPRD,siva,Cao:2021mwx,Dadhich:2022yuk,hogan,barrow1,ong2,abo,Ong:2018xna,barrow2,barrow3,b1,MYNEWPRD,Ong:2018xna,Gurzadyan:2021hgh,DiGennaro:2021fak,GRF,loeb2022three}. %
In 2002, Gary Gibbons and, independently, Schiller deduced the factor 1/4 and showed that force at a point is never larger that the maximum value $c^4/4G \approx 3.0 \cdot 10^{43} \rm \,N$ \cite{gibbons,s1}. 
The maximum value is realized on black hole horizons.
At that time, it also became clear that the field equations of general relativity and the 
Hilbert action can be \emph{deduced} from the invariant maximum force ${c^4}/{4G}$ \cite{s1,csmax,siva,MYNEWPRD,GRF}. %

The maximum force value $c^4/4G$ is due to the maximum energy per
      distance ratio appearing in general relativity.
Indeed, for a Schwarzschild black hole, the ratio between its energy $Mc^2$ 
      and its diameter $D=4GM/c^2$ is given by the maximum
      force value, independently of the size and mass of the black hole. 
Also the force on a test mass that is lowered with a rope towards a gravitational 
      horizon -- whether charged, rotating or both -- 
      never exceeds the force limit, but only when the minimum size of the test mass is taken into account.  
All apparent counter-examples to maximum force disappear when explored in detail \cite{jowsey,faraoni,newcs,Jowsey:2021gny,Faraoni:2021sep,MYNEWPRD,Volovik:2022sdr}.

A maximum force  implies that space is curved. 
The maximum force value is realized at horizons.
%
In fact, maximum force $c^4/4G$ \emph{implies} Einstein's field equations of general relativity.
There are at least two ways to deduce the field equations from maximum force \cite{s1,csmax,siva,MYNEWPRD,GRF}. %
Maximum force also implies the cosmological constant term, but does not fix its value.
As a consequence, the maximum force limit can be seen as the defining \emph{principle} of general relativity.
The situation resembles special relativity, of which the maximum speed limit
can be seen as the defining principle. %
The invariant limit property of $c^4/4G$ thus goes beyond a conversion factor
between curvature and energy density. 

Because maximum force implies general relativity with the cosmological constant, 
also the usual big-bang cosmology follows from maximum force. 
Maximum force implies all observed aspects of gravitation.

The maximum force principle for general relativity is not the only possible principle.
Other maximum quantities combining $c$ and $G$, such as maximum power 
$c^5/4G$ \cite{sciama1,MYNEWPRD,lastsivaram,massa,hogan,Kostro:2000gw,maxlum,Ong:2018xna,abo,Gurzadyan:2021hgh} 
or maximum mass flow rate $c^3/4G$ \cite{MYNEWPRD,Cao:2021mwx}, can also be taken as principles of relativistic gravity. %
%
\textcolor{black}{Also the length to mass limit $c^2/4G$, realized by black holes, can be taken as defining general relativity.}
\textcolor{black}{Each of these equivalent limits can be taken as starting principle of general relativity.
Maximum force is chosen here only because it is the most striking of these limits.}

Attempts to find counterexamples to maximum force (or the other 
equivalent limits) are not successful. 
In flat space and at low speeds, the maximum force value implies inverse square gravity \cite{cshoopmax}, which is well established experimentally. %
Because the force limit is \emph{local}, an observer cannot add forces acting at different location and claim that their sum exceeds the local limit $c^4/4G$. %
(Such examples are easily found.) 
The value $c^4/4G$ is also the largest possible gravitational force between two black holes. %
Maximum force also implies the hoop conjecture \cite{thorne,Hod:2020yub,Liu:2021fit,cshoopmax}.
Furthermore, maximum force eliminates most, but not all, alternative theories of gravity \cite{MYNEWPRD}.
However, it is unclear whether modified Newtonian dynamics remains possible or is eliminated.

No counterexample to the maximum luminosity and power value $c^5/4G \approx 9 \cdot 10^{51} \rm \,W$ has been found. 
Even the most recent observations of black hole mergers fail to exceed the luminosity limit; %
the highest instantaneous luminosity observed so far is about 0.5\% of the maximum value.
Also in cosmology, no power value exceeding the limit is observed \cite{MYNEWPRD,loeb2022three}. 

Falsification of the limits is possible. It is sufficient to observe or to point out a value for local force, power or luminosity that exceeds the respective limit. %
The probability is low.
Every day, maximum force and general relativity are confirmed by the position determination performed by mobile phones with satellites.

\textcolor{black}{Again, it is important to note that the principle of maximum force, together with the principle of maximum speed and the principle of least (or stationary) action, implies and contains diffeomorphism invariance.}

In short, general relativity can be deduced \textcolor{black}{from the 
principle of least (or stationary action), 
the \emph{principle of maximum speed}, and} the
\emph{principle of maximum force:}
\begin{equation}
	F \leq c^4/4G \;\;.
\end{equation}
There is a force limit in nature.
More precisely, the Hilbert action, Einstein's field equations of general relativity, and diffeomorphism invariance can be 
deduced from the {principle of maximum force} combined with the {principle of maximum speed} and 
the principle of least action. %
The principle of maximum force was the last building block that allowed summarizing physics in 9 simple lines.

\section{The quantum of action} 


\noindent
Quantum theory is based on the invariant smallest action $\hbar \approx 1.1 \cdot 10^{-34} \rm \,Js$.
It is not possible to measure action values -- i.e., changes -- smaller than $\hbar$, a constant of nature that is called the elementary quantum of action. 
(In fact, the smallest change is $h=2 \pi \hbar$, but often the two quantities are used interchangeably.)
The quantum of action is the origin of the indeterminacy relation. 
Above all, the quantum of action explains photons and atoms.

Planck discovered the quantum of action $\hbar$ in the 1890s, when studying light. 
The term `quantum' was introduced by Galileo, who explained that matter is made of `piccolissimi quanti', tiny quanta, that are not divisible. %
In 1906, following Einstein, Planck took over the term \cite{planck1906vorlesungen}. 

In nature, action is \emph{quantized}.
An action value, or change, smaller than $\hbar$ is never measured \cite{cohen2019quantum,bartelmann2018theoretische,zagoskin2015quantum,levy1998quantique}. 
In addition, every action value -- every measured change -- is a multiple of $\hbar$.
This property also implies the quantization of angular momentum. 
The quantum of action is so fundamental that it is used to define the kilogram in the international system of units.
The limit $\hbar$ requires to introduce wave functions, Hilbert spaces and operators. 
This leads to the Schrödinger equation, the Dirac equation and all of quantum theory, including probabilities and entanglement \cite{cohen2019quantum}. %
The quantum of action $\hbar$ modifies the principle of least action in the microscopic domain:
it determines the mathematical structure of Lagrangians using operators and quantum states that correctly describe the probabilistic outcomes of experiments \cite{cohen2019quantum,bartelmann2018theoretische,zagoskin2015quantum}.

A straightforward attempt to falsify smallest action is to measure a system's or a particle's 
energy $E$ twice, once at the start and once at the end of an interval $\delta t$.
Even though in the classical approximation action is given by the product $W=E \,\delta t$ and can get as small as desired, in nature -- and in quantum theory -- the action value $W$ remains finite when $\delta t$ gets small:
the measured energy (difference) increases when $\delta t$ decreases. 
The reason is the uncertainty relation: it prevents that the measured action value approaches zero when $\delta t$ does so.

Other attempts at finding a counter-example to the quantum of action use spin. 
Because action is quantized in multiples of $\hbar$, there is no spin smaller than 1/2:
detecting a spin 1/2 flip requires an action $\hbar$. 
There is no way to detect a spin flip with a smaller amount of action.

A further attempt is the detection of light. 
But even detecting even the dimmest light requires an action $\hbar$. 
Light consists of photons.
In nature, there is no way to detect one half or one hundredth of a photon. 
Photons are elementary quanta: they cannot be split.
If $\hbar$ were not the smallest action value, photons would not exist.
Also atoms would not exist without the lower limit set by $\hbar$.
The invariant limit property of $\hbar$ thus goes beyond a conversion factor between angular velocity and energy, or between wave number and momentum.

Action quantization is confirmed by all experiments ever performed.
The discovery of $\hbar$ led to the development of electronics, lasers, computers and the internet.
Indeed, no (non-equivalent) alternative description of quantum physics has ever been proposed.
Nevertheless, falsification remains possible, by measuring a smaller action value than the quantum of action $\hbar$. %
It is unlikely that this will happen.

In short, combining the principle of least action with the \textit{quantum of action}
\begin{equation}
	W \geq \hbar  
\end{equation}
implies quantum theory. 
In line with the above statements one can state: quantum theory can be deduced from the \emph{principle of quantized action.} 
The quantum of action implies the Lagrangian of quantum theory. 
In particular, when the speed limit $c$ is included into quantum theory, antiparticles, the Dirac equation and quantum field theory arise. %

\section{The Boltzmann constant}


\noindent
Whether thermodynamics is part of fundamental physics or not has been a subject of debate.
Cohen-Tannoudji, Okun, and Oriti are among those in favour \cite{COHENT,Okun:2001rd,Oriti:2018tym}.
Therefore, it is included here.

Classical thermodynamics can be seen, to a large extent, as a consequence of the principle of least action.
Similarly, statistical physics can be seen as following from quantum theory.
Indeed, there are uncertainty relations for thermodynamic properties. 
As an example, temperature $T$ and energy $U$ obey $\Delta (1/T) \; \Delta U \geq {k}/{2}$.
This relation was first given by {Bohr}; it was discussed by {Heisenberg} and other scholars \cite{Uffink1999,Shalyt-Margolin:2003qig,Hasegawa:2022czt}.
It suggests that entropy is similar to action, with the Boltzmann constant $k$ times 
$\mathcal{O}(1)$ 
taking the role of $\hbar$.

Planck introduced and named the Boltzmann constant $k \approx 1.4 \cdot 10^{-23} \rm \,J/K$ together with $\hbar$.
Is $k$ a just unit conversion factor between energy and temperature or is it related to a fundamental limit?
In 1929, Szilard suggested \cite{szilard} that there is a smallest entropy in nature. 
Since then, the concept of a `quantum of entropy' has been explored by many authors \cite{BRILL,zi1,zi2,zi3,zimo1,zimo2,zimo3,zimo4,zimo5,garcia1993quantum,COHENT,Liao_2004,kirwan2004intrinsic,Meschke2006,Kothawala:2008in,Liu_2009,ren2010entropy,Dydyshka:2013eza,BAKSHI2017334,Yu-Qi,JQQ,PhysRevB.93.155404,schwab1,schwab2,schwab3,Partanen2016,ferenc}.
Entropy is observed to be quantized in various systems: 
in electromagnetic radiation \cite{kirwan2004intrinsic,Meschke2006}, in the entropy of two-dimensional electron gases \cite{PhysRevB.93.155404} and in low temperature thermal conductance \cite{schwab1,schwab2,schwab3,Partanen2016,ferenc}. 
These investigations conclude that there is a smallest entropy value, which is given by a multiple of $k$.
Often, but not always, the smallest entropy is given as $k\,\ln 2$, as done by Szilard. 
In modern terms, this numerical factor expresses that the smallest possible entropy is related to a single bit.

The concept of a smallest entropy was explored in detail by Zimmermann \cite{zimo1,zimo2,zimo3,zimo4,zimo5}
and by Lavenda \cite{LAVENDA}. 
They deduced statistical mechanics from the existence of such a smallest entropy value in nature. 
The invariant limit property of the smallest entropy thus goes beyond a conversion factor
between temperature and energy. 
(As a note, combining statistical mechanics with quantum theory yields and explains decoherence.)

Entropy quantization is confirmed by all experiments ever performed.
Every time a thermometer is read out and every time hot air rises, the relevance of the Boltzmann constant is confirmed.
Nevertheless, falsification is possible, by measuring a smaller value than the quantum of entropy. %
Also in this case, it is unlikely that this will happen.

It has to be stressed that the quantum of entropy does \emph{not} imply a smallest value for the entropy \emph{per particle}, but a smallest entropy value for a physical system.
For interacting systems of particles, entropy values \emph{per particle} can be much lower than the limit. 
In Bose-Einstein condensates, measured values for the entropy per particle can be as low as $0.001 k$ \cite{naturek}. 

In short, there is a smallest entropy value in nature.
Continuing the above collection of limits, one can state: statistical thermodynamics can be deduced from 
\begin{equation}
	S \geq k \ln 2  \;\;.
\end{equation}
%
This is the \emph{principle of smallest entropy}.

\section{Electromagnetism}


\noindent
The theory of quantum electrodynamics is based on the U(1) gauge symmetry \textcolor{black}{(or U(1) gauge invariance)} of electromagnetism.
The gauge symmetry determines the (minimal) coupling of the Dirac equation to the electromagnetic field.
The vector potential in the Dirac equation has a local phase freedom that is called \emph{gauge} freedom \cite{Pauli:1941zz}. %
The U(1) gauge group explains the vanishing mass of the photon, Coulomb's law, magnetism and light.
When particle properties (of line 8) are included, U(1) implies charge conservation, Maxwell's equations \cite{heras2007can,burns2019maxwell}, stimulated emission, Feynman diagrams, and perturbative quantum electrodynamics. %
This in turn yields the change or `running' of the fine structure constant and of the electron mass, as well as all other observations in the domain, without any exception. %

The description provided by quantum electrodynamics and the corresponding experiments match to high precision.
Deviations between calculation and experiments are possible, but have not been found yet.
Clever measurement set-ups for the well-known $g$-factor of the electron yield results with 13 to 14 significant digits that all agree with calculations \cite{Gabrielse:2019cgf}. %
Even in the case of the muon $g$-factor, there is still no confirmed deviation between experiment and calculation \cite{muon1,Muong-2:2021ojo}. %
In everyday life, every laser confirms quantum electrodynamics.

In short, combining least action, the quantum of action, maximum speed and the 
\begin{equation}
	\text{U(1) gauge group}
\end{equation}
with the particle properties and the fine structure constant of line 8 and 9 below, fully specifies and describes electromagnetism, both in the quantum and the macroscopic domain. %
The Dirac equation for charged particles and the Lagrangian of {QED} arise in this way.
For example, the {QED} Lagrangian explains all observed material properties.

\section{The nuclear interactions}


\noindent
The strong and the weak nuclear interactions are based on an SU(3) and a broken SU(2) gauge symmetry \textcolor{black}{-- or on the corresponding gauge invariances (broken in the case of the weak interaction).}
They define strong charge and weak charge, as well as all their properties and effects.
For example, the gauge groups explain the burning of the Sun, radioactivity, and the history of the atomic nuclei found on Earth. %

The verification of the two non-Abelian gauge theories -- with all their detailed particle properties, particle reactions, and consequences for nuclear physics -- took many decades \cite{pdgnew}. %
The verification was completed when accelerator experiments confirmed the existence of the Higgs boson in 2012.
Both gauge groups also imply the running of the fundamental constants with energy.
Attempts at falsification or even just at extension of the gauge description -- such as the search for a fifth force, grand unification, more gauge bosons, etc. -- were not successful, despite intense research all over the world \cite{pdgnew}. %
Also the recent W boson mass measurement is not a confirmed deviation \cite{CDF:2022hxs}.

In short, the combination of least action, the quantum of action, the speed limit and the gauge groups
\begin{equation}
	\text{SU(3) and broken SU(2)}   
\end{equation}
fully specifies and describes the nuclear interactions, including the Lagrangians of QCD and of the weak interaction, provided the particle spectrum and the fundamental constants given in the following are included. %

\section{The particle spectrum}


\noindent
The world around us is made of elementary fermions and bosons.
All matter consists of fermions: six types of quarks and six types of leptons.
All radiation is made of gauge bosons -- the photon, the W, the Z and gluons -- and of the predicted graviton.
The Higgs boson, giving mass to all particles, completes the list.
The Higgs boson also explains the breaking of SU(2) gauge symmetry.

Each elementary particle is described by mass, spin, electric charge, weak charge, colour charge, parities, baryon number, lepton number and the flavour quantum numbers. %
No other 
particle property has been detected.
All the known particle properties and their conservation laws have been explored in great detail. 
Every two years, the Particle Data Group documents the status and experimental progress across the world \cite{pdgnew}. %

In short, everything observed is made of  
\begin{equation}
	\text{18 elementary particles.}   
\end{equation}
Nature specifies these particles and their properties.
One can also speak of 18 fundamental fields.
The particle number 18 arises if all gluons are counted as one particle, and if the coloured quarks and all the antiparticles are not counted separately. %
The essence of the statement is that the 18 fermions and bosons just mentioned suffice to build everything observed in nature, and that they \textit{fix} the full mathematical expression for the Lagrangian of the standard model -- together with the last line. %
Therefore, these elementary particles and their properties need to 
appear in Table~1.

\section{The fundamental constants}


\noindent 
The standard model is specified with 25 characterizing numbers. They include 15 elementary particle masses (or more precisely, the ratios to the Planck mass), 3 coupling constants, as well as 6 mixing angles and 2 CP phases both in the CKM (Cabibbo-Kobayashi-Maskawa) and in the PMNS (Pontecorvo-Maki-Nakagawa-Sakata) mixing matrices for quarks and neutrinos \cite{pdgnew}. %
One parameter is redundant. 
Because the constants run with energy, the precise statement is that the standard model is described by 25 fundamental constants at some defined energy value. 
Two further characterizing numbers, the cosmological constant and the number of spatial dimensions, determine the expansion of space-time. %
In accordance with all present experiments, nature is thus described by 27 fundamental constants.
Together, these 27 specific numerical values determine the remaining details of the Hilbert Lagrangian and of the standard model Lagrangian. %

The last fundamental constants in the standard model Lagrangian have been introduced in the 1970s. 
All the values are being measured with a precision that usually increases when new experiments are performed \cite{pdgnew}.
At present, the fundamental properties of the neutrinos are the least precisely known.
The cosmological constant in the Lagrangian of general relativity has been introduced more than a century ago. 
After a complicated history, its value was first measured in the 1990s.

Neither general relativity nor the standard model \textit{explain} the values of the fundamental constants.
Explaining these values -- which include the mass of the electron and the fine structure constant $1/137.036(1)$ -- remains an open issue. %
These two particular constants almost completely determine the colours in nature.
As long as the numbers are unexplained, colours are not fully understood.

Various attempts to reduce the number of fundamental constants have been proposed.
Most attempts predict new effects that have not been observed.
Other proposals, such as certain kinds of supersymmetry, require additional fundamental constants.
However, no additional fundamental constant has yet been discovered \cite{pdgnew}. %

In short, nature somehow chooses
\begin{equation}
	\text{27 fundamental constants}   
\end{equation}
that, together with the previous lines, completely determine the Hilbert Lagrangian of general relativity and the Lagrangian of the standard model of particle physics (with massive neutrinos). %

\section{The summary of present physics}   

\noindent 
Lines 1, 2, 3 and 9 fully determine the Hilbert Lagrangian, including the cosmological constant.
The derivation is found in references \cite{MYNEWPRD} and \cite{GRF}.
Line 5 determines thermodynamics, as shown in references \cite{zimo1,zimo2,zimo3,zimo4,zimo5} and \cite{LAVENDA}. 
All lines except 3 and 5 fully determine the Lagrangian of the standard model of particle physics: they determine  
the elementary particle spectrum, the particle mixing matrices, the particle masses, their couplings, the interaction terms, 
the kinetic terms, and, as a result, the full Lagrangian.
The complete expression of the Lagrangian of the standard model is derived in references \cite{diagrammatica} 
(for vanishing neutrino mass) and \cite{pdgnew} (for massive neutrinos). %
The corresponding lines in Table~1 have exactly the same physical and mathematical content, while avoiding  
writing down the algebraic details that are implied by them. %

While the number of lines in Table~1 is subjective, the content is not.
The number could easily be expanded or reduced by one or two lines, while keeping the same content.
Whatever form is chosen, the content of the lines in Table~1 agrees with all experiments.
Only standard textbook physics is included. 
No part of standard textbook physics is missing.

Table~1 resulted from the work of many thousands of scientists and 
engineers during over 400 years. %
Galileo started around the year 1600, with the first-ever measurements of the dynamics of moving bodies. 
Line 1, the principle of least action, was fully formulated around 1750. 
Line 5, on thermodynamics, arose from 1824 to 1929, and line 6, on electrodynamics, arose around 1860.
Line 2, on maximum speed came around 1890, and line 4, about the quantum of action, around 1900.
Line 3, on maximum force, was implicitly given in the year 1915, and formulated in 2002.
As a result, the Hilbert Lagrangian of general relativity agrees with experiments since more than 100 years.
The remaining lines 7 to 9, on the standard model, arose in the years from 1936 to 1973.
The standard model Lagrangian of particle physics thus agrees with experiments since about 50 years.

Given that no observation contradicts the two Lagrangians and 
   thermodynamics, one can say that the 9~lines contain 
   all present knowledge about nature, including all textbook physics and all observations ever made. %
The 9~lines also contain chemistry, material science, biology, medicine, geology, astronomy and engineering. %
This is the conclusion of a world-wide and decade-long effort to evaluate the 9~lines.
The simplicity of the 9~lines and their vast domain of validity form an intriguing contrast. 

The 9~lines contain five general principles and four lines of specific choices taken from an infinity of possibilities.
The five principles define the framework of modern physics.
The four lines of choices specify the everyday world and, at the same time, contain what is unexplained about modern physics.

In short, the 9~lines of Table~1, five principles and four sets of choices, 
contain the evolution equations of the standard model, of general relativity and of thermodynamics.
The 9~lines describe all of nature, without any deviation between theory and experiment.
\textcolor{black}{This controversial} summary leads to several questions, \textcolor{black}{challenges} and predictions.

\textcolor{black}{\section{What advantage do the 9 lines provide?}}

\noindent
The formulation of physics given in Table~1 is compact, but provides no new content. Is it still useful?
It turns out that the 9 lines do imply a shift in perception about several important issues in relativistic quantum gravity.

It is regularly suggested that general relativity and quantum theory are incompatible.
Often, the incompatibility is even called a contradiction. %
Table~1 suggests that this is not the case, 
and that, instead, the principles \textit{complement} each other.
The five principles appear to address different aspects of nature, and all of them in a similar way.
Indeed, experiments have never found a contradiction between general relativity and the standard model.
In fact, such contradictions only arise when extrapolations into inaccessible domains are made.
This conclusion can be stated with precision.

Combining the experimental limits on speed $v$, force $F$ and action $W$ using 
the general relation for energy $E=Fvt=W/t$ leads to a limit on  
measurements of time $t$ given by %
\begin{equation}
  t \geq\sqrt{{4G\hbar}/{c^5}} \approx 1.1 \cdot 10^{-43}{\,\rm s} \;\;.
\end{equation}
The five principles thus \emph{eliminate instants of time} and introduce a minimum time interval, given by twice the Planck time.
In the same way, the principles also \emph{eliminate points in space} and introduce a minimum length \cite{mead}
given by 
\begin{equation}
  l \geq\sqrt{{4G\hbar}/{c^3}} \approx 3.2 \cdot 10^{-35}{\,\rm m} \;\;,
\end{equation}
twice the Planck length.
As a consequence, continuity of space and time is not intrinsic to nature, but due to an averaging process. 

\textcolor{black}{Conversely, the assumption of fully continuous space implies a vanishing Planck length. This implies a vanishing quantum of action or a vanishing gravitational constant, or both.}
Full continuity thus implies the lack of most physical measurement units.
In other words, if continuity is taken as an \textit{exact} property of nature, it contradicts modern physics.

All issues disappear if space and time are seen as (effectively) continuous \textit{only} for all intervals \textit{larger} than the Planck limits. 
The existence of a smallest length and of a smallest time interval -- and of all Planck limits in general -- 
ensures that \textit{quantum theory and general relativity never actually contradict each other.}
No accessible domain of nature yields a contradiction between the two theories. 
This conclusion is confirmed by studies from particle physics, such as reference \cite{ELLIS2009369}.

\textcolor{black}{A smallest observable length value also implies that there is no way to ever detect or measure additional spatial dimensions -- be they microscopic or macroscopic.}
In other words, \textit{higher spatial dimensions cannot arise in nature.}
Indeed, there are no experimental hints for such additional dimensions.

The minimum length also implies that space has no scale symmetry, no conformal symmetry, and no twistor structure. 
The length measurement limit also prevents the observation and the existence of lower dimensions. 
A smallest observable length further implies, together with the observed isotropy and boost invariance, that space is neither discrete nor a lattice. 
The length measurement limit prevents the observation and the existence of any additional spatial structure at the Planck scale, including non-commutativity or additional symmetries.

In short, the 9~lines imply that continuity is approximate, that space has three curved dimensions without additional structure, that the minimum length eliminates most past unification attempts, and that general relativity and the standard model are not contradictory or mutually exclusive, but that they \textit{complement} each other.
%

\section{Are there more than 9 lines?}


\noindent 
Candidates for disagreement between the 9-line summary and experiment arise regularly.
Examples are W mass measurements, muon $g-2$ measurements, dark energy differing from the cosmological constant, dark matter, the rotation curves of galaxies, or table-top quantum gravity. %
It could be that a future experiment will require changes in the 9~lines. 
\textcolor{black}{(It has to be stressed that elementary dark matter particles have not been detected yet.)}
Therefore, these and other candidates for disagreement are being explored around the world in great detail. %
Even though no confirmed observation is unexplained by the 9~lines, the \emph{experimental quest} for such 
an effect will never be over.

In particular, all the experiments that confirmed lines 6 to 9 make a further statement.
The specific choices contained in these four lines imply that additional interactions, 
additional particles, or additional constants would greatly increase the complexity of the 
table -- in contrast to observations.

Are unexplained observations possible at all? 
In other terms, are \textit{additional lines} necessary to describe nature? 
The simplicity and consistency of the 9~lines suggest a negative answer.
So far, proposals for physics beyond the standard model either require more lines, like supersymmetry, or, if they don't, like grand unified theories, they disagree with experiment \cite{shifman2019musings}. 
Some proposals, like additional elementary dark matter particles, require more lines \textit{and} disagree with experiment.
Nevertheless, a future unexplained observation \textit{cannot} be excluded.


In short, the 9~lines suggest the lack of new physics. 
Any observation or experiment unexplained by the 9~lines will create a sensation \textcolor{black}{and would falsify most of the conclusions presented in this article.}

\section{What experimental predictions follow?}

\noindent 
The 9~lines summarizing physics can and should be tested in as many future \textit{experiments} as possible. 

\textbf{Prediction~1.} Lines 2 to 5 imply that \emph{no trans-Planckian quantities or effects} arise in nature.
This is valid for length, time and for every other physical observable.
The precise limits are given by the \textit{(corrected) Planck limits}, such as the minimum length $\sqrt{4 G \hbar/ c^3}$, the minimum time $\sqrt{4 G \hbar/ c^5}$ or the maximum force $c^4/4G$. %
Here, the factor 4 from maximum force corrects the commonly used Planck units.
(For some cases, such as Planck energy or Planck momentum, the derivation of the limit is only valid for a single elementary particle.) %
For example, a limit on length measurements also implies that no experiment will observe singularities, discrete space-time, additional space-time structures, or additional dimensions. %
%
%
These predictions are in agreement with all experiments. 
They also confirm the prediction of the lack of infinitely large and of infinitely small observables made by Hilbert in 1935 \cite{hilbert1935naturerkennen}.

\textbf{Prediction~2.} A continuous space-time in spite of the existence of a minimum length implies that \emph{locality, continuity, causality and three-dimensionality are valid at all observable scales}, 
i.e., at all scales larger than the Planck scale. %
The lack of trans-Planckian effects prevents the observability, the influence and the existence of other dimensions or other structures in space-time \textcolor{black}{-- both microscopic and macroscopic.}
This prediction agrees with all data so far.

\textbf{Preditction~3.} The 9~lines predict that there is \emph{no physics beyond special relativity, beyond general relativity, beyond thermodynamics, beyond quantum theory and beyond the standard model.} %
The nine lines predict the lack of any additional symmetries, structures or effects whatsoever, \emph{at any length or energy scale:} 
     the lines predict the so-called \emph{high-energy desert.} %
In the past centuries, mistaken predictions about the lack of new physics have been made several times.
At present however, there is a difference: the prediction agrees with all high-precision observations since over five decades. %

In short, the nine lines predict a lack of new physics.
Interestingly, this prediction does not completely exclude the observation of \textit{new} relativistic quantum gravity effects. 
(The lines 6 to 9 can be seen as \textit{known} relativistic quantum gravity effects.)
Also the observation of non-relativistic quantum gravity effects remain possible \cite{hossenfelder}.
However, if any such new effects are detected, they will not contradict Table~1.

\section{Are there fewer than 9 lines?}

\noindent 
Each of the 9~lines in Table~1 generates a question about its origin. 
In particular, one can ask for the origin of the five \emph{principles} listed in the lines 1 to 5.
So far, no accepted explanation for the origin of the principle of least action nor for the other limit principles has been proposed. 
It is unclear how nature enforces its five principles.
In modern terms, it is unclear how the principles \textit{emerge} from an underlying description.

The lack of explanation is especially evident in lines 6 to 9.
These four lines contain all the \emph{specific choices} that fix the details of the standard model and of general relativity. %
At present, the origins of the force and particle spectra are unknown, as is the origin of each fundamental constant. %
One can say that so far, these four lines are the only known observations \emph{beyond} the standard model and \emph{beyond} general relativity. %
However, despite multiple and intense efforts, no explanation for the four lines of {specific choices} has been successful and accepted. %
The mechanism of their emergence is unclear.

The lack of explanations despite the successful description of nature with Table~1 leads to a related question: 
     can the 9~lines be deduced from a smaller set? %
All these queries are part of the other, \emph{theoretical quest} being pursued in fundamental physics.

The five principles of lines 1 to 5 are not good candidates to 
shorten Table~1, because they are independent of each other. %
It appears impossible to reduce the number of principles in lines 1 to 5 in a simple way. 
Only a radical explanation based on emergence might have a chance to reduce their number.

In contrast, reducing the number of specific choices given in lines 6 to 9 should be possible.
The specific choices out of an apparent infinity of options are so particular that they cannot be fundamental.
The four lines \emph{must} be due to a deeper, emergent explanation.
In the past five decades, various proposals to reduce the number of
lines 6 to 9 -- or simply their details -- have been made. 
However, no proposal agrees with observations to full precision. 

In short, a description of nature with less than 9~lines must exist.
Such a description must be emergent.
So far, it has not been found. 
Nevertheless, Table~1 yields several suggestions.

\textcolor{black}{\section{Implications for the microscopic constituents of nature}}

\noindent 
The 9~lines provide clear hints for any description of nature based on emergence. The 9~lines describe the behaviour of curvature, radiation and matters, thus of space and particles.

\comment{
The five principles use time and space but forbid the existence of points and instants.
There are no continuous sets in nature.
Continuity is a macroscopic approximation.
But the fundamental measurement limits also prevent the existence of sharp boundaries and thus of discrete sets in nature.
In nature, sets only arise after approximations or idealizations.
All axioms are based on sets.
Without sets, axioms cannot be formulated.
Therefore, the five principles \textit{prevent} an axiomatic description of nature. 
Despite Hilbert's dream of an axiomatic description of physics, formulated in his sixth problem, the five principles only allow a consistent and (logically) circular description of nature.
}

In any complete description of nature, space and particles must emerge from some common description. 
The horizon of a black hole makes this especially clear: it can bee seen as due to collapsing matter and can be seen as curved space.
The common description of matter and space must be based on {common microscopic degrees of freedom, or microscopic constituents}.
These microscopic constituents describe extended space and curvature, localized particles, probabilistic quantum motion, and the minimum length.

To realize the minimum length, the constituents must be of minimum size, i.e., of Planck size, in at least one dimension.
To realize the macroscopic extension of space, the constituents must have \emph{at least one} macroscopic dimension.
To reproduce probabilities, particles must be made of fluctuating constituents.
To allow both fluctuations, particle localization and particle motion, the constituents must be of codimension two: they must have \emph{at most one} macroscopic dimension.
As a result, the fluctuating constituents must resemble thin fluctuating strands of macroscopic length and Planck-size radius.
In other terms, \emph{the constituents of space and particles must be fili\-form, with Planck radius.}

More precisely, the microscopic constituents of space and particles differ from points in two ways: %
   they are \emph{fili\-form} and they are \emph{discrete}, i.e., countable.
Their discreteness implies and confirms 
the \emph{finiteness} of black hole entropy, the Bekenstein entropy bound \cite{entropybound} and the maximum entropy emission rate \cite{mirza}. 
In fact, their discreteness 
implies all Planck limits.
Their fili\-form structure explains, as shown by Dirac's trick \cite{gardner}, the spin 1/2 behaviour of fermions, i.e., their different behaviour under rotations by $2 \pi$ and by $4 \pi$.
The fili\-form structure also explains the surface-dependence of black hole entropy 
\cite{weber}.


\textcolor{black}{In short, the 9~lines of Table~1 imply that the description of space and of the known particles must emerge from \textit{fluctuating} constituents that are \emph{fili\-form} and \textit{have Planck radius}.} 
This conclusion agrees with decades-old arguments and allows a number of predictions on how to achieve an even shorter summary of physics.
\bigskip

\textcolor{black}{\section{Predictions about the theory of relativistic quantum gravity}}

\noindent 
The 9~lines and the discrete fili\-form constituents of space and matter imply a number of theoretical predictions.

\textbf{Prediction~4.} It was shown above that the limits $c$, $c^4/4G$, $\hbar$ and $k\,\ln 2$  define special relativity, general relativity, quantum theory and thermodynamics.
In the same way, the limits arising when combining those theories -- i.e., when combining the four limits, such a minimum length or minimum time -- \emph{define} the complete theory of relativistic quantum gravity. %
More precisely, the 9~lines imply that relativistic quantum gravity is already 
known \emph{in all its experimental and theoretical effects:} 
relativistic quantum gravity implies the five principles of line~1 to~5
and it fixes the choices of lines~6 to~9. %

\comment{
\textbf{Pr.~6.} The limits $c$, $c^4/4G$, $\hbar$ and $k\,\ln 2$ lead to  minimum length and time intervals.
They imply that space and time are \emph{not} made of points.
Likewise, point particles do \emph{not} exist, and interactions are not perfectly local.
%
%
%
The microscopic constituents of matter and space that are at the basis of the limits to locality are discrete and fili\-form.
%
%
In other terms, \emph{continuity emerges from averaging microscopic fluctuating constituents.}
}

\textbf{Prediction~5.} 
Continuity emerges from averaging fluctuating fili\-form constituents.
This must apply to space, to fields, and to wave functions.
Because continuity arises in all settings that allow measurements -- and despite the existence of a smallest length and time intervals -- space and time \emph{can} be used to describe nature. %
In fact, because all the limits $c$, $c^4/4G$, $\hbar$, and $k\,\ln 2$ 
contain meter and second in their units, space and time \emph{must} be used to describe nature. %
There is no chance to find a back-ground-free description of motion, nor to describe nature without time.
The limits imply that relativistic quantum gravity must continue to use -- as is done by general relativity and by the standard model --  locally one-dimensional time and locally three-dimensional space.
The use of space and time remains necessary despite the impossibility to define distances and
time intervals below the Planck limits.

\textbf{Prediction~6.} 
In contrast to the situation in special relativity, in general 
relativity and in quantum theory, no simple physical system \emph{realizes} any limit of relativistic quantum gravity -- i.e., any limit containing $c$, $G$ and $\hbar$ but not the Boltzmann constant $k$.
Light moves with $c$, electron spin flips allow to measure $\hbar$, black hole horizons realize $c^4/4G$.
In contrast, the microscopic constituents of nature are out of experimental reach: no simple physical system realizes the Planck length, the Planck energy, the Planck time or any other limit of relativistic quantum gravity.
As a result, \emph{no evolution equation} for the microscopic constituents of relativistic quantum gravity can be defined.
Such an equation would not be testable.

Equations in relativistic quantum gravity can only be found for large numbers of microscopic constituents, such as in black holes, where the smallest length arises in the expressions for the entropy or for the temperature, and the Boltzmann constant $k$ is therefore used.
In relativistic quantum gravity, only \emph{statistical effects} can be expected to be calculated or to be observed.  
Apart from black hole entropy, the Bekenstein bound and the Unruh effect are examples of calculations. 
Curvature, wave functions and particle masses are examples of observations that are due to large numbers of microscopic constituents.
In contrast, single microscopic constituents cannot be described nor observed.

In other terms, \emph{there will never be an evolution equation or a Lagrangian for relativistic quantum gravity.} 
This result requires a change in thinking habits.
Evolution equations and Lagrangians have been seen as essential part of physics for over four centuries.
In stark contrast, the complete theory of relativistic quantum gravity is \emph{defined completely} by the limits $c$, $c^4/4G$, $\hbar$, $k \ln 2$ and by the statistics of the microscopic constituents -- without additional or new equations of motion.
The conclusion confirms the statement that no Lagrangian can explain the appearance of Lagrangians -- in the same way that turtles cannot support turtles all the way down.

In short, Table~1 implies that in the complete theory of relativistic quantum gravity, all nine lines follow from \textit{large numbers} of \textit{fili\-form constituents} of \textit{Planck radius} that \textit{fluctuate in $3+1$ dimensions.} No equations of motion can describe the constituents. Only a statistical description of their behaviour is possible. This result confirms the expectations for any emerging and complete description of nature. The lack of equations of motion also explains the limited use of mathematical formulae in this article.
\bigskip

\textcolor{black}{\section{Other approaches}}

\noindent
The theoretical predictions about the complete theory of relativistic quantum gravity that were just given are unusual.
Many approaches differ.

A number of past approaches use concepts that contradict the smallest length.
Examples are approaches that use conformal symmetry, scaling symmetry, non-commutative space, higher or lower spatial dimensions, space-time foam, vortices, twistors,
tetrahedra, or additional group structures. 

A large part of past approaches contradict the smallest time value: this applies to all approaches  based on Lagrangians.
A further group of approaches attempts to avoid space-time completely and explores background-independent descriptions.

Other past approaches use different microscopic constituents, such as strings, membranes, loops, spins, sets, lines, triangles, or bands.

Various past approaches to relativistic quantum gravity predict new observations in particle physics, such as new energy scales, new particles, new symmetries, new interactions and reactions, or new fundamental constants. 
Other approaches predict deviations from general relativity  \cite{addazi2022quantum}. 

However, none of these approaches fully explains the four lines of choices in Table~1, despite intense efforts.
This also holds for proposals that \textit{extend} the content of Table~1. At the same time, none of these approaches is based on fluctuating filiform constituents with Planck radius. 

In short, the approaches to a complete theory of relativistic quantum gravity that were explored in the past appear less promising than the direction suggested by the 9~lines summarizing physics.
\bigskip

\section{Towards a theory of relativistic quantum gravity}

\noindent
The specific choices in lines 6 to 9 must \emph{emerge} from the theory of relativistic quantum gravity. %
This is a demanding requirement.

So far, the requirement is \emph{not fully realized} by any proposed kind of microscopic constituents. %
In other terms, the statistical behaviour of fluctuating microscopic 
constituents will and must explain 
the gauge groups, 
the spectrum of elementary particles, their mass values, their mixing angles and CP phases, the values of 
the coupling constants, the cosmological constant and the number of spatial dimensions. %

As long as Table~1 remains valid, checks of lines 6 to 9 are also the \emph{only tests possible at 
present} for the correctness of any proposed model of microscopic constituents. %
In fact, counting the lines of Table~1 that are explained and tested 
successfully by a given research approach is a practical way to quantify its achievements.
Among the required tests, explaining the last line of Table~1, line 9, is the decisive one.

Line 9 specifies the masses of elementary particles (or, equivalently, the mass ratios to the Planck mass), 
the coupling constants and the mixing matrices. 
These pure numbers are fundamental constants that describe the world around us.
These fundamental constants are not yet explained by any research approach that also agrees with the first 5 lines.
A way to calculate the fundamental constants needs to be found.
(References  \cite{cspepan,csorigin,csindian,csqed,botta,carlip} explore a possible starting point.)
In addition, line 9 specifies the number of dimensions and the cosmological constant. 
Also these two numbers describe the world around us and must be explained.

In short, in the search for relativistic quantum gravity, the most productive way forward appears to be the following. 
   First, propose a specific microscopic model of space and matter that is based on fili\-form constituents of Planck radius and that realizes the five principles. 
   Afterwards, check its consequences for the four choices of lines 6 to 9. 
In the decisive test, the fundamental constants need to be calculated and compared with experiment.
\bigskip

\section{Conclusion and outlook}  

\noindent 
Present physics -- experiment and theory -- can be condensed in 9~lines that
   describe all observations to full precision and determine both the Lagrangian of general relativity and the Lagrangian of the standard model. %
The 9 lines consist of the five principles of least action, of maximum speed, of 
   maximum force, of action quantization and of smallest entropy, plus  four lines of specific choices for the gauge interactions, the elementary particles, and the fundamental constants. %

The main experimental prediction of the 9~lines is the lack of any effect beyond general relativity and beyond the standard model of particle physics, with massive neutrinos. %
The main theoretical prediction is that some discrete, fili\-form and fluctuating constituents of Planck radius in $3+1$ dimensions will explain all 9~lines, not with an equation of motion or a Lagrangian, but instead only with statistical arguments.

\textcolor{black}{The complete theory of relativistic quantum gravity -- the ``theory of everything" or ``final theory" -- is predicted to have only two experimental effects:
general relativity and
the standard model of elementary particle physics, with massive neutrinos.}
Therefore, the decisive test of any proposed complete theory is the calculation of the values of the elementary particle masses, of the mixing matrices and of the coupling constants.
\bigskip

\section*{Acknowledgments and declarations}  
\label{sec:ack}

\noindent 
The author thanks Chandra Sivaram, Arun Kenath, Erik Baigar, Lucas Burns, Thomas Racey, Michael Good, Peter Woit, Louis Kauffman, Michel Talagrand, Luca Bombelli, Isabella Borgogelli Avveduti, Peter Schiller, Steven Carlip and \textcolor{black}{an anonymous referee} for discussions. %
Part of this work was supported by a grant of the Klaus Tschira Foundation.
The author declares that he has no conflict of interest and no competing interests.
There is no additional data available for this manuscript (if all experiments ever made are put aside).

\comment{
L. Szilard \ti
\"{U}ber die Entropieverminderung in einem thermodynamischen System bei
Eingriffen intelligenter Wesen/ \jo Zeitschrift f\"{u}r Physik/ \vo 53/ \pp
840-856/ \yrend 1929/ English translations: _Behavioral
Science_, v. 9, pp. 301-310 (1964); B. T. Feld and G. Weiss Szilard,
_The Collected Works of Leo Szilard: Scientific Papers_, (MIT Press,
Cambridge, 1972), pp. 103-129; J. A. Wheeler and W. H. Zurek,
_Quantum Theory and Measurement_ (Princeton University Press), pp.
539-548; and Leff and Rex (1990), pp. 124-133.
}

\bibliography{9linesu}

\end{document}